# Systematic investigation of the deformation mechanisms of a γ-TiAl single crystal

Byungkwan Jeong[a], Jaemin Kim[a], Taegu Lee[a], Seong-Woong Kim[b,**], Seunghwa Ryu[a,*]

[a] Department of Mechanical Engineering & KI for the NanoCentury, Korea Advanced Institute of Science and Technology, Daejeon 34141, Republic of Korea

[b] Titanium Department, Korea Institute of Materials Science, Changwon 51508, Republic of Korea

[*] Corresponding author. Tel.: +82-42-350-3019; fax: +82-42-350-3059

Email address: ryush@kaist.ac.kr (Seunghwa Ryu)

[**] Co-corresponding author. Tel.: +82-55-280-3837; fax: +82-55-280-3255

Email address: mrbass@kims.re.kr (Seong-Woong Kim)

## Abstract

We propose a theoretical framework to predict the deformation mechanism of the γ-TiAl single crystal without lattice defects by combining the generalized stacking fault energy and the Schmid factor. Our theory is validated against an excellent testbed, the single crystal nanowire, by correctly predicting four major deformation mechanisms, namely, ordinary slip, super slip, twinning, and mixed slip/fracture observed during the tensile and compressive tests along 10 different orientations using molecular dynamics simulations. Interestingly, although lattice defects are not taken into account, the theoretical predictions match well with existing experiments on bulk specimen with only a few exceptions; the exceptions are discussed based on the size-dependent deformation mechanism in the presence of preexisting dislocation sources. We expect that the method in this paper can be generalized to study various ductile intermetallic crystals where conventional Schmid law does not hold well.

## 1. Introduction

Over the past few decades, TiAl intermetallic compound has attracted much attention because of its use in applications such as LPT (low pressure turbine) blades in aircraft engines and turbine wheel in automobile engines enabled by its outstanding high temperature mechanical properties including good thermal and oxidation resistance[1-4]. Despite these excellent properties, industrial applications have been limited by its inherently low ductility. To develop the compound with enhanced ductility, the microscopic deformation mechanisms must be understood[1].

TiAl alloy is composed of the brittle $\alpha_2$ phase and the ductile $\gamma$ phase, and its plasticity is accommodated mostly by the ductile $\gamma$ phase. Numerous studies have been performed to understand the mechanical behaviors of TiAl alloy and each of the constituent phases[5-14]. The brittleness of $\alpha_2$ has been explained by the limited number of active slip systems, based on the detailed analysis on its slip systems and dislocation structures[7,10,14]. Deformation mechanisms of single-crystal $\gamma$-TiAl, prepared either from TiAl alloy or by direct growth for various orientations have been investigated via transmission electron microscopy (TEM) or scanning electron microscopy (SEM) analyses[5,11,13]. The effects of different $\gamma/\gamma$ interfaces and lamellae orientations of TiAl compound have also been studied[6,12]. In addition, the motion of existing dislocations and their effects on mechanical properties in $\gamma$-TiAl have been investigated[15,16,17].

However, although its unique slip system (Fig. 1) including both ordinary dislocations and superdislocations has been extensively studied[18,19], a systematic prediction on the deformation behaviors of γ-TiAl is still lacking, thereby limiting the understanding of the mechanical properties of TiAl alloy such as orientation-dependent yield stress, elongation, and deformation mode of PST (polysythetically twinned) crystals[20]. Although a few analyses based on the Schmid factor exist[5,18], it is well known that the Schmid factor alone is not sufficient to predict the deformation mechanism of single-crystal metals[21-23]. For example, in an experimental study[5], γ-TiAl single crystals under compression along seven loading directions at root temperature undergo plastic deformation mediated by ordinary dislocations except one direction, although the Schmid law predicts that only two loading directions prefer the ordinary dislocation while the superdislocation or the twinning mechanism has the highest Schmid factor for the other five directions. In addition, although both superdislocations and ordinary dislocations are decomposed into partial dislocations, existing studies consider the Schmid factors for full ordinary dislocations and full superdislocations without accounting for the partial slips[5,18].

In the present study, in order to overcome the limited applicability of the Schmid law in the γ-TiAl crystal, we suggest a theory to predict the deformation mode of the crystal without lattice defects by combining generalized stacking fault energy (GSFE) surface and the Schmid factor accounting for the partial slips. We test the theory against an excellent testbed, the single crystal nanowire, by performing molecular dynamics (MD) simulations of both compressive and tensile tests along 10 different orientations. We reveal that four major deformation mechanisms of ordinary slip (slip by ordinary dislocation), super slip (slip by superdislocation), twinning, and mixed slip/fracture occur depending on the loading condition, and that the theoretical prediction matches with the deformation mode observed in MD simulations for all loading conditions tested in the study. Having established the validity of our theory, we predict the deformation mode of single crystal compression experiments by employing the GSFE obtained from first-principle density function theory (DFT) calculations, and show that the predictions match well with experimental results in the literature[5,13,24] with a few exceptions. We discuss the origin of the discrepancy by considering size-dependent deformation mechanism in the presence of preexisting dislocation sources.

The remainder of this paper is organized as follows. In Section 2, we discuss the unique slip systems in detail and present a theoretical model to predict the preferred deformation mode for a given loading condition. We calculate the ideal critical resolved shear stress (ICRSS) of five distinct slip events (forward super slip, SISF (superlattice intrinsic stacking faults) partial slip, twin, inverse super slip, and ordinary slip) and provide a systematic method to compute the critical stress for each to occur based on the GSFE surface and the Schmid factor. In Section 3, we provide a detailed description of the MD simulations and show that the observed deformation mechanisms for 20 different loading conditions match well with the theoretical predictions. We then discuss the origin of different deformation mechanisms between the theory and the existing bulk experiments for a few exceptional orientations by considering size-dependent deformation mechanism in the presence of preexisting dislocation sources. A summary and an outlook for future research are presented in Section 4.

## 2. Theoretical Predictions on the Deformation Mechanism
### 2.1. Characterization of Slip Systems

γ-TiAl single crystal has an $L1_0$ intermetallic structure, which is composed of alternating Ti and Al atomic layer along the [001] axis, as shown in Fig. 1(a). γ-TiAl single crystal has almost the same atomic configuration with the face-centered cubic (FCC) structure, except that the lattice constant along the [001] axis is approximately 2% longer than the lattice constant along the [100] or [010] axes. To distinguish two distinct orientations, we use the modified notation of Miller indices with mixed parentheses of $\langle abc]$ or $\{def)$ introduced by Hug et al.[25]. The perfect $L1_0$ crystal structure can also be described as the repeated stacking of three consecutive $\{111)$ atomic layers. A stacking fault is a planar defect, where these stacking sequences are interrupted by the relative displacement between two adjacent planes. This interruption involves excess energies compared to a perfect crystal, called the generalized stacking fault energy (GSFE). The GSFE surface is calculated by moving upper half of the perfect crystal along the directions parallel to the $\{111)$ plane (or gamma surface) and calculating the energy difference between the initial state and the deformed state. The GSFE landscape of the γ-TiAl single crystal is different from that of the FCC structure, as shown in Figs. 1(c) and (d), and three different stacking faults can form that include superlattice intrinsic stacking faults (SISF), antiphase boundary (APB) and complex stacking faults (CSF). The properties of the γ-TiAl single crystal, including lattice constant, elastic constants and three stacking fault energies (SISF, APB and CSF) obtained from two EAM potentials (Farkas & Jones[26] and Zope & Mishin[27]), *ab initio* density functional theory (DFT) calculations and experiments are summarized in Table 1.

Because of the aforementioned unique GSFE, the γ-TiAl crystals deform differently from the FCC crystals. According to previous experimental studies[5,11,13], plastic deformation of bulk single-crystal γ-TiAl involves twinning, ordinary dislocation and superdislocation on the $\{111)$ plane. Among these, the formation of superdislocation is a unique feature that distinguishes γ-TiAl from ordinary FCC structures. Two types of superdislocations with burgers vectors of ⟨011] and 1/2⟨112] are often considered, as shown in Fig. 2. According to the generalized Peierls-Nabarro model based on the *ab initio* GSFE surface[19], fully dissociated superdislocations involve four partials and three stacking faults, as follows:

$$[01\bar{1}] = 1/6[11\bar{2}] + SISF + 1/6[\bar{1}2\bar{1}] + APB + 1/6[11\bar{2}] + CSF + 1/6[\bar{1}2\bar{1}] \quad (1)$$

$$1/2[11\bar{2}] = 1/6[11\bar{2}] + SISF + 1/6[\bar{1}2\bar{1}] + APB + 1/6[11\bar{2}] + CSF + 1/6[2\bar{1}\bar{1}] \quad (2)$$

Because of the limited experimental resolution, many experimental studies reported that superdislocations dissociate into three partial dislocations and two stacking faults [25,28] or two partials and SISF[13,29,30]. Nevertheless, in the present theoretical study, we considered the fully dissociated superdislocations with four partials and ordinary dislocation with two partials.

The slip on the {111) plane is initiated either by one SISF partial or two CSF partials, as shown in Fig. 3(a). On one hand, when a SISF partial is formed in the first stage, three possible events can subsequently occur (Figs. 3(b), (d), and (e)). First, three trailing partials nucleate on the same slip plane of the leading SISF partial, leading to the super slip. We name this mechanism as forward super slip to distinguish another super slip mechanism whose leading partial is the CSF partial. Second, another leading partial can nucleate on a parallel plane not adjacent to the first slip plane, leading to SISF partial slip, i.e., a formation of multiple SISFs on different slip planes. Third, another leading partial nucleation occurs on the slip plane adjacent to the first leading partial, leading to twinning. On the other hand, when a CSF partial is formed in the first stage, two possible events of either ordinary slip or super slip can occur, depending on the following trailing partials (Fig. 3(c)). We name this super slip as the inverse super slip because the leading partial of a superdislocation is usually considered to be SISF partial rather than the CSF partial because of the lower energy barrier to form SISF. In total, we considered five possible deformation events. For the inverse super slip, the ⟨011] superdislocation is preferred to the 1/2⟨112] superdislocation because of a geometrical reason. After forming a CSF partial slip and then another slip from CSF to APB, the third partial slip direction determines which of two superdislocations forms (Fig. 3(a),(c)). Whether upward or downward CSF partial slip occurs in the first step (Fig. 3(a)), the third partial slip direction of the ⟨011] superdislocation (Fig. 3(c)) is aligned with the first CSF partial direction, while the third partial slip direction of the 1/2⟨112] superdislocation is aligned with the other CSF partial direction that is not chosen in the first step. Since the CSF partial with higher Schmid factor is chosen in the first step, the third partial of the ⟨011] superdislocation has higher Schmid factor than that of the 1/2⟨112] superdislocation. Because of the similar geometrical reason, the ⟨011] superdislocation is preferred to the 1/2⟨112] superdislocation for the forward super slip. Hence, we only consider the ⟨011] superdislocation in the present study. In addition, because the CSF or APB fault has higher energy compared to SISF, we do not consider the events of forming multiple CSF or APB faults on different slip planes.

We note that the GSFE surface moderately changes with the strain[31]. Because the pre-strain originated from the surface stress of nanowire is non-negligible and the applied strain at the yield point is even higher, it may be necessary to incorporate the strain-induced change of GSFE surface to enhance the accuracy of deformation mode prediction. However, as shown in the later part, although we use the GSFE obtained at zero strain for the prediction of the deformation mechanism, we find that our theoretical predictions match very well with MD simulations for all 20 loading conditions considered in the study. Indeed, the previous study on FCC crystals[23] also shows that the deformation mode can be predicted accurately without accounting for the strain-induced GSFE change.

**2.2 Prediction of the Deformation Mechanism.**

For each partial dislocation involving the aforementioned five categorized events, we obtain the GSFE curve by projecting the GSFE surface along the partial Burgers vector directions constituting each event, as shown in Fig. 4. Because the plastic deformation of the nanowire initiates from the dislocation nucleation in the absence of preexisting dislocations, we compute the ideal critical resolved shear stress $\sigma^{ICRSS}$ of each partial slip from the maximum slope of the corresponding section of GSFE curve, as summarized in Table 2. Although the $\sigma^{ICRSS}$ calculation in this work assumes the sliding of the perfect half space sliding instead of dislocation nucleation from a surface, it will determine the preference between different slip events; a similar method has been applied to predict the deformation mode of FCC crystals by Cai & Weinberger[23].

We extended the methodology to the intermetallic system involving more complex GSFE surface with more deformation modes. We note that the lattice friction resistance of both $L1_0$ crystal and FCC crystal is not significant because the dislocation core structure of both crystals involves stacking faults and thus is planar. Although the core structure of the $1/2\langle110]$ ordinary screw dislocation in TiAl gamma crystal is known to be non-planar exceptionally, the lattice friction stress (the Peierls stress) at 0K is estimated by a DFT method to be relatively small value of around $0.01C_{44}$ ($C_{44}$=68 GPa)[32]. The lattice friction stress at 300K is expected to be significantly smaller due to thermal fluctuation[33]. Hence, although the lattice friction resistance is not taken into account, the deformation mode prediction based on the minimum crystal stress criterion works well for FCC crystals [23] and also for $L1_0$ crystal of the TiAl alloy in this study. We note that our method is not expected to work at a very low temperature where the lattice resistance of the ordinary screw dislocation becomes significant.

For a given axial (tensile or compressive) loading, we can obtain the Schmid factor $S$ of each partial slip and calculate the critical axial stress by $\sigma^c = \sigma^{ICRSS}/S$. We first compare the critical axial stresses of one SISF partial and two CSF partials for all available {111) slip planes and choose the preferred slip direction with lower critical stress. We note that, because we only consider three directions toward saddle points of the GSFE surface in the Step-1 deformation (Fig. 3a), negative Schmid factors appear in our analysis (footnote of Table 2), which is different from the conventional usage of Schmid law on ordinary dislocations in FCC crystals. We do not consider any slip event along the slip directions with negative Schmid factors due to their larger energy barriers. When the SISF direction is preferred in the first step, we compare the critical stresses of three possible events of partial slip $\sigma_{SISF}^c$, twinning $\sigma_{twin}^c$, and forward super slip $\sigma_{Fsuper}^c$. Although the superdislocation forms by three trailing partials, we set $\sigma_{Fsuper}^c = \sigma_{SISF\ to\ APB}^c$ because the other two partials, one from APB to CSF and the other from CSF to perfect crystal, have significantly lower ICRSS values than those of the two previous partials, $\sigma_{SISF}^{ICRSS}$ and $\sigma_{SISF\ to\ APB}^{ICRSS}$, respectively. Hence, the critical stress of forward super slip is defined as that of the partial slip from SISF to APB. The preferred deformation mechanism can be visualized in a map spanned by two parameters, $\tau_1 = \sigma_{SISF}^C/\sigma_{twin}^C$ and $\tau_2 = \sigma_{Fsuper}^C/\sigma_{twin}^C$, as shown in Fig. 5(a). We note that twinning is always preferred than the multiple SISF formations because $\sigma_{twin}^{ICRSS}$ is lower than the $\sigma_{SISF}^{ICRSS}$ and their Schmid factors are identical. Alternatively, when the CSF partial is preferred in the first step, we compare the critical stresses of two possible events of ordinary slip $\sigma_o^c$ and inverse super slip $\sigma_{Isuper}^c = \sigma_{CSF\ to\ APB}^c$. We visualized the preferred mechanism along the $\tau_3 = \sigma_o^c/\sigma_{Isuper}^c$ axis, as depicted in Fig. 5(b). Table 3 summarizes the deformation mode prediction for 20 loading conditions based on the theoretical framework combining GSFE and the Schmid factor.

3. **Comparison with Molecular Dynamics Simulations and Experiments**

## 3.1. Molecular Dynamics Simulations and GSFE calculations

We use LAMMPS (large-scale atomic/molecular massively parallel simulator) software [34] to investigate the deformation mode of a single-crystal nanowire by MD simulations. First, we construct γ-$Ti_{50}Al_{50}$ nanowires with 3:1 aspect ratio to have 18 nm height and 6 nm diameter with given orientation. The periodic boundary condition is imposed only along the loading direction (x-axis). Prior to the loading tests, we equilibrate the specimen at 300 K by using NPT ensemble simulation for 30 ps, during which the equilibrium length is obtained. We further equilibrate the nanowire by 30 ps of NVT ensemble simulation at the equilibrium length. Next, we perform uniaxial tension and compression tests of nanowires along the x-axis with $10^8 s^{-1}$ strain rate up to 30% strain by imposing incremental strain of 0.1% every 10 ps with 300 K NVT ensemble. To understand the orientation dependent deformation mechanism, we perform the uniaxial compression and tension tests for 10 different crystallographic orientations: ⟨001], ⟨010], ⟨011], ⟨102], ⟨110], ⟨111], ⟨112], ⟨120], ⟨201], and ⟨211]. The simulation results are visualized using OVITO (the open visualization tool)[35] and POV-Ray[36].

The choice of a reasonable interatomic interaction model is crucial for the classical MD simulations. We test two embedded-atom method (EAM) interatomic potentials developed by Farkas & Jones[26] and Zope & Mishin[27]. A key difference between two EAM interatomic potentials is the relative magnitude of the APB and CSF energies in the relaxed GSFE map. As depicted in Figs. 6(d) and (e), the APB energy is higher than the CSF energy for the Farkas & Jones model, whereas the order is opposite for the Zope & Mishin model. In comparison, in Fig. 6(f), the density functional theory (DFT) calculations show that the APB energy is higher than the CSF energy and that there is no minimum at the APB. Although both EAM models do not match quantitatively with the DFT result, the Farkas & Jones model is better suited for studying the deformation mechanism, which is sensitive to the landscape of the GSFE map. Hence, we use the Farkas & Jones EAM for describing the interatomic interactions.

We also explain the details about the GSFE landscape in Fig. 6(c) and (e) calculation based on density functional theory (DFT) using VASP (Vienna *Ab initio* Simulation Package)[37]. The exchange-correlation functional was described within the Generalized gradient approximation (GGA) and parameterized by Perdew, Burke, and Ernzerhof[38]. K-point meshes were constructed using the Monkhorst-Pack scheme[39]. We used a $11 \times 5 \times 19$ k-point mesh and a plane-wave cutoff energy of 320 eV. To calculate the stacking fault energy, we constructed 6 atomic layers along (111) plane, having spacing of 13.8 Å, 4.9 Å and 2.8 Å along the (111) slip plane normal and $[11\bar{2}]$ and $[\bar{1}10]$ slip directions, respectively. In-plane displacement was imposed along the $[11\bar{2}]$ and $[\bar{1}10]$ directions and accompanied by relaxation along the (111) axis. From the calculation, APB stacking fault has no stable state, and its energy is higher than CSF.

### 3.2. Comparison between Theoretical Predictions and MD Simulations

In a series of MD simulations, we found that γ-TiAl nanowires deform by four types of mechanisms: ordinary slip, twin, super slip and mixed slip/fracture mode. In order to identify four deformation modes, we calculated relative displacement vectors of the atoms above and below of the slip planes. This vector was calculated by the position of the atoms for each step when they are moved. Stacking faults were identified by common-neighbor analysis. Four representative deformation mechanisms are visualized in Fig. 7, and detailed simulation results for all loading conditions are summarized in Fig. S2 of Supplementary Information. For the twinned nanowires, a twin nucleates at the yield point and expands linearly with tensile loading by forming new partial dislocations on adjacent slip planes of the existing twin. Ordinary slips or super slips occurred on one or multiple {111) planes. We find that all observed super slips in the present work occurred via the inverse super slip mechanism. The specimen deformed in the mixed slip/fracture mode showed irregular deformation behaviors involving cracks and ⟨011] superdislocations formed by inverse super slip. When inverse super slip is predicted from the theoretical framework, we observe the super slip mechanism (Fig. 7(d)) when the critical stress is relatively small. However, when the critical stress for inverse super slip is high, small crack-like deformation also occurs simultaneously, leading to the mixed slip/fracture (Fig. 7(c)).

For all loading conditions considered in the study, we find a good match between the theoretical prediction and simulation results, as shown in Table 3. When the CSF partial is preferred to the SISF partial, the ICRSS of the trailing partial of ordinary dislocation is significantly lower than the ICRSS of the trailing partials of superdislocation. Hence, ordinary slip is preferred to the inverse slip for the majority of loading conditions. However, we observe the superdislocation by inverse super slip in a few loading conditions. For the mixed slip/fracture case involving inverse super slip, the Schmid factor of trailing partial of the ordinary dislocation is equal to or less than zero. When the SISF partial is preferred to CSF partial in the first step, the critical stress of twin is always lower than those of multiple SISFs and forward super slips for all loading conditions tested in the study.

Our analysis also provides a qualitative understanding of the stress-strain curves of different loading conditions, as shown in Fig. 7(e). The different yield stresses (or failure stresses) in MD simulations can be understood based on the critical stress (Table 3). Although the realistic slip occurs via the dislocation loop nucleation from the surface, we assume rigid crystal block sliding when estimating the critical stress. Hence, the critical stress predictions in Table 3 (9.73 GPa for ⟨110]T, 9.43 GPa for ⟨011]T, 21.96 GPa for ⟨001]T, 11.01 GPa for ⟨102]T) overestimates the MD results. Nevertheless, the yield stress in MD simulations reflects the critical stress prediction to some extent, although the order of yield stress can be slightly different because of the nonlinear stress-strain curve and the detailed dislocation nucleation mechanism of each.

### 3.3. Comparison between Theoretical Predictions and Existing Experiments

Having justified the validity of the theoretical framework by comparing its prediction with MD simulation results, we compared the results from our combined theoretical and computational study with existing experiments. Interestingly, for the ⟨001] compression condition in which twinning is observed in our nanowire simulations, experiments on bulk single crystal reported the formation of superdislocations[5,13] in a wide range of temperature. To make a more appropriate comparison with experiments, we predicted the preferred deformation modes based on the ICRSS from DFT calculations (Table 4) and compared them with existing compression experiments along 9 orientations[5,13,24], as shown in Table 5. Although most experiments consider a non-stoichiometric crystal and bulk specimen which inherently have preexisting lattice defects, our results match well with experiments, only except for the [001], [$\bar{1}$52] and [$\bar{2}$33] orientations. In addition to the difference in chemical composition, we discuss the plausible origin of discrepancy in two aspects: the error in GSFE prediction from DFT and the operation of different deformation mechanism due to pre-existing defects in macroscale specimen. One can also suspect the surface stress effect originating from high surface to volume ratio, but we have shown that the effect of free surface on the deformation behavior is rather limited.

We first note that the critical stress predictions for ordinary slip and inverse super slip differ by only 11% for [$\bar{1}52$] orientation, as shown in detail in the Table S1 of the Supplementary Information. Different atomic composition, the temperature effect, and the limitation of DFT functional accuracy can be the sources of the different deformation mechanisms. Interestingly, two loading conditions, [001] and [$\bar{2}33$] compressions, for which twinning is predicted by our theoretical framework, are reported to be deformed by superdislocations. The conclusion remains valid when we also test a new twinning path reported recently by Wang et al[40], as depicted in the Fig S3 of the Supplementary Information. Because our GSFE-based model cannot explain such discrepancy, we tried to understand the observation based on the size dependent preferred deformation modes in the presence of preexisting defects, as described in the previous studies[41-44].

The typical size of the specimen in the experiment[5,24] (a few millimeters) is considerably larger than the nanowires used in MD simulation; therefore, the specimen generally contains defects that can act as the Frank-Read source, pre-existing statistically stored dislocations, and etc. In comparison, the nanowire in our MD simulation is a lattice-defect-free specimen whose deformation is governed by dislocation nucleation. The critical stress criterion based on the Schmid factor and GSFE are relevant to predict the nucleation of dislocation or twin in defect-free nanocrystals. However, as the sample size increases, it becomes progressively important to consider the stress to form superdislocation and twin in the entire domain in the presence of pre-existing defects, as discussed in the literature[41-44].

The resolved shear stress required to emit a superdislocation in the presence of a Frank-Read source can be expressed as follows:

$$\tau_s/\mu_s = b_s/d \qquad (5)$$

where $\tau_s$ is the critical resolved shear stress to emit a $[01\bar{1}]$ superdislocation on the (111) plane, $b_s$ is the magnitude of the Burgers vector of the superdislocation, $\mu_s$ is the effective shear modulus for superdislocation, and $d$ is the sample size (or grain size for polycrystal). Fig. 8(a) shows that this equation is derived from the equilibrium relation of line tension of superdislocation and resolved shear stress applied to superdislocation. In addition, the resolved shear stress required to emit a twinning partial can be expressed as follows:

$$\tau_p/\mu_p = b_p/d + \gamma_{SISF}/(\mu_p b_p) \qquad (6)$$

where $\tau_p$ is the critical resolve shear stress to emit a $1/6[11\bar{2}]$ twinning partial on the (111) plane, $b_p$ is the magnitude of the Burgers vector of the twinning partial, $\mu_p$ is the effective shear modulus for the twinning partial, and $\gamma_{SISF}$ is the stacking fault energy. Details of the derivation of (6) can be found in the Supplementary Note 1. Unlike the case of superdislocation, we must consider the effect of stacking fault energy for the generation of the twinning partial (Fig. 8(b)). A transition from superdislocation to twinning will occur as the size of the specimen is decreased below a critical grain size (which can be considered as the maximum size of Frank-Read source), where the critical stress for emitting the twinning partial becomes equal to that for superdislocation. The critical grain size ($d_c$) can be obtained by equating (5) and (6) and is given as follows:

$$d_c = b_p(\mu_s b_s - \mu_p b_p)/\gamma_{SISF} \qquad (7)$$

The critical size is predicted to be 48.9 nm from the Farkas & Jones EAM potential ($b_p=0.1616$ nm, $b=0.5579$ nm, $\gamma_{SISF}=87.51$ mJ/m², $\mu_s=72.05$ GPa, and $\mu_p=85.01$ GPa) and 18.8 nm from the first principle calculation ($b_p=0.1647$ nm, $b=0.5685$ nm, $\gamma_{SISF}=179.1$ mJ/m², $\mu_s=53.76$ GPa, and $\mu_p=61.39$ GPa). The parameters required for the calculation are summarized in Table 1, and the effective shear modulus was calculated based on the method of Scattergood and Bacon (see the Supplementary Note 3 of the Supplementary Information)[45]. Hence, we could suspect that the deformation twinning is unlikely to occur in the experimentally-tested millimeter scale single crystal samples which are orders of magnitude larger than the critical size predicted from theoretical calculations.

## 4. Conclusion

In summary, we suggest a theoretical framework to analyze the deformation behavior of a γ-TiAl single crystal without lattice defects and benchmarked the theory against MD simulations. MD simulations revealed that the γ-TiAl single-crystal was deformed by four types of deformation behaviors: ordinary slip, twinning, super slip, and mixed slip/fracture. We predict the deformation mode based on the critical stress calculation obtained from the Schmid factor and the GSFE curve and find a good match with the simulation results. Hence, we show that the limited applicability of the Schmid law on the γ-TiAl can be resolved when GSFE curve is also taken into account. Interestingly, our theoretical predictions on the deformation mechanism matched well with existing experiments on bulk specimen with only a few exceptions, although the bulk specimen inherently have preexisting lattice defects. We suggest size-dependent deformation mechanism originated from the preexisting dislocation source as a plausible cause of the discrepancy, in addition to the inaccuracy of the DFT calculations and different chemical composition. We note that the suggested analysis method in the present study can be applied to other intermetallic systems involving the formation of superdislocations.

# References

1. Appel, F. *et al.* Recent progress in the development of gamma titanium aluminide alloys. *Adv Eng Mater* **2**, 699-720, doi:Doi 10.1002/1527-2648(200011)2:11<699::Aid-Adem699>3.0.Co;2-J (2000).

2. Clemens, H. & Mayer, S. Design, Processing, Microstructure, Properties, and Applications of Advanced Intermetallic TiAl Alloys. *Adv Eng Mater* **15**, 191-215, doi:10.1002/adem.201200231 (2013).

3. Dimiduk, D. M. Gamma titanium aluminide alloys - an assessment within the competition of aerospace structural materials. *Mat Sci Eng a-Struct* **263**, 281-288, doi: 10.1016/S0921-5093(98)01158-7 (1999).

4. Kim, Y. W. Ordered Intermetallic Alloys .3. Gamma-Titanium Aluminides. *Jom-J Min Met Mat S* **46**, 30-39 (1994).

5. Inui, H., Matsumuro, M., Wu, D. H. & Yamaguchi, M. Temperature dependence of yield stress, deformation mode and deformation structure in single crystals of TiAl (Ti-56 at% Al). *Philos Mag A* **75**, 395-423, doi: 10.1080/01418619708205149 (1997).

6. Kanani, M., Hartmaier, A. & Janisch, R. Stacking fault based analysis of shear mechanisms at interfaces in lamellar TiAl alloys. *Acta Mater* **106**, 208-218, doi:10.1016/j.actamat.2015.11.047 (2016).

7. Kerans, R. J. Deformation in Ti3al Fatigued at Room and Elevated-Temperatures. *Metall Trans A* **15**, 1721-1729, doi: 10.1007/Bf02666355 (1984).

8. Kim, S. W., Na, Y. S., Yeom, J. T., Kim, S. E. & Choi, Y. S. An in-situ transmission electron microscopy study on room temperature ductility of TiAl alloys with fully lamellar microstructure. *Mat Sci Eng a-Struct* **589**, 140-145, doi:10.1016/j.msea.2013.09.080 (2014).

9. Kim, S. W., Wang, P., Oh, M. H., Wee, D. M. & Kumar, K. S. Mechanical properties of Si- and C-doped directionally solidified TiAl-Nb alloys. *Intermetallics* **12**, 499-509, doi:10.1016/j.intermet.2004.01.004 (2004).

10. Lipsitt, H. A., Shechtman, D. & Schafrik, R. E. The Deformation and Fracture of Ti3al at Elevated-Temperatures. *Metall Trans A* **11**, 1369-1375, doi: 10.1007/Bf02653491 (1980).

11. Mahapatra, R., Girshick, A., Pope, D. P. & Vitek, V. Deformation Mechanisms of near-Stoichiometric Single-Phase Tial Single-Crystals - a Combined Experimental and Atomistic Modeling Study. *Scripta Metall Mater* **33**, 1921-1927, doi: 10.1016/0956-716x(95)00460-D (1995).

12. Palomares-Garcia, A. J., Perez-Prado, M. T. & Molina-Aldareguia, J. M. Effect of lamellar orientation on the strength and operating deformation mechanisms of fully lamellar TiAl alloys determined by micropillar compression. *Acta Mater* **123**, 102-114, doi:10.1016/j.actamat.2016.10.034 (2017).

13. Stucke, M. A., Vasudevan, V. K. & Dimiduk, D. M. Deformation-Behavior of [001]Ti-56al

Single-Crystals. *Mat Sci Eng a-Struct* **192**, 111-119, doi: 10.1016/0921-5093(94)03224-6 (1995).

14  Wiezorek, J. M. K., Court, S. A. & Humphreys, C. J. On the dissociation of prism plane superdislocations in Ti3Al. *Philosophical Magazine Letters* **72**, 393-403, doi:10.1080/09500839508242479 (1995).

15  Simmons, J. P., Rao, S. I. & Dimiduk, D. M. Atomistics simulations of structures and properties of 1/2 <110> dislocations using three different embedded-atom method potentials fit to gamma-TiAl. *Philos Mag A* **75**, 1299-1328, doi: 10.1080/01418619708209858 (1997).

16  Katzarov, I. H., Cawkwell, M. J., Paxton, A. T. & Finnis, M. W. Atomistic study of ordinary (1)/(2) < 110] screw dislocations in single-phase and lamellar gamma-TiAl. *Philos Mag* **87**, 1795-1809, doi:10.1080/14786430601080252 (2007).

17  Katzarov, I. H. & Paxton, A. T. Atomistic studies of < 101] screw dislocation core structures and glide in gamma-TiAl. *Philos Mag* **89**, 1731-1750, doi:10.1080/14786430903037281 (2009).

18  Appel, F., Paul, J. D. H. & Oehring, M. *Gamma titanium aluminide alloys: science and technology*.  (John Wiley & Sons, 2011).

19  Schoeck, G., Ehmann, J. & Fahnle, M. Planar dissociations of [101] superdislocations in TiAl: ab-initio electron theory and generalized Peierls-Nabarro model. *Philosophical Magazine Letters* **78**, 289-295, doi: 10.1080/095008398177869 (1998).

20  Inui, H., Oh, M. H., Nakamura, A. & Yamaguchi, M. Room-temperature tensile deformation of polysynthetically twinned (PST) crystals of TiAl. *Acta Metallurgica et Materialia* **40**, 3095-3104, doi:10.1016/0956-7151(92)90472-Q (1992).

21  Inui, H., Nakamura, A., Oh, M. H. & Yamaguchi, M. Deformation Structures in Ti-Rich Tial Polysynthetically Twinned Crystals. *Philos Mag A* **66**, 557-573, doi: 10.1080/01418619208201575 (1992).

22  Kishida, K., Inui, H. & Yamaguchi, M. Deformation of lamellar structure in TiAl-Ti(3)Al two-phase alloys. *Philos Mag A* **78**, 1-28, doi: 10.1080/01418619808244799 (1998).

23  Weinberger, C. R. & Cai, W. Plasticity of metal nanowires. *J Mater Chem* **22**, 3277-3292, doi:10.1039/c2jm13682a (2012).

24  Feng, Q. & Whang, S. H. Cross-slip and glide behavior of ordinary dislocations in single crystal gamma-Ti-56Al. *Intermetallics* **7**, 971-979, doi: 10.1016/S0966-9795(99)00005-9 (1999).

25  Hug, G., Loiseau, A. & Veyssiere, P. Weak-Beam Observation of a Dissociation Transition in Tial. *Philos Mag A* **57**, 499-523, doi:Doi 10.1080/01418618808204682 (1988).

26  Farkas, D. & Jones, C. Interatomic potentials for ternary Nb-Ti-Al alloys. *Model Simul Mater Sc* **4**, 23-32, doi: 10.1088/0965-0393/4/1/004 (1996).

27  Zope, R. R. & Mishin, Y. Interatomic potentials for atomistic simulations of the Ti-Al system. *Phys Rev B* **68**, doi:10.1103/PhysRevB.68.024102 (2003).

28  Court, S. A., Vasudevan, V. K. & Fraser, H. L. Deformation Mechanisms in the Intermetallic Compound Tial. *Philos Mag A* **61**, 141-158, doi: 10.1080/01418619008235562 (1990).

29  Gregori, F. & Veyssiere, P. Properties of < 011]{111} slip in Al-rich gamma-TiAl I. Dissociation, locking and decomposition of < 011] dislocations at room temperature. *Philos Mag A* **80**, 2913-2932, doi: 10.1080/01418610008223902 (2000).

30  Hug, G., Loiseau, A. & Lasalmonie, A. Nature and Dissociation of the Dislocations in Tial Deformed at Room-Temperature. *Philos Mag A* **54**, 47-65, doi: 10.1080/01418618608242882 (1986).

31  Aubry, S., Kang, K., Ryu, S. & Cai, W. Energy barrier for homogeneous dislocation nucleation: Comparing atomistic and continuum models. *Scripta Mater* **64**, 1043-1046, doi:10.1016/j.scriptamat.2011.02.023 (2011).

32  Woodward, C. & Rao, S. I. Ab-initio simulation of (a/2)< 110] screw dislocations in gamma-TiAl. *Philos Mag* **84**, 401-413, doi:10.1080/14786430310001611626 (2004).

33  Kang, K., Bulatov, V. V. & Cai, W. Singular orientations and faceted motion of dislocations in body-centered cubic crystals. *Proceedings of the National Academy of Sciences* **109**, 15174, doi: 10.1073/pnas.1206079109 (2012).

34  Plimpton, S. Fast Parallel Algorithms for Short-Range Molecular-Dynamics. *J Comput Phys* **117**, 1-19, doi: 10.1006/jcph.1995.1039 (1995).

35  Stukowski, A. Visualization and analysis of atomistic simulation data with OVITO-the Open Visualization Tool. *Model Simul Mater Sc* **18**, doi:10.1088/0965-0393/18/1/015012 (2010).

36  Cason, C. J., Froehlich, T., Kopp, N. & Parker, R. POV-Ray for Windows. *Persistence of Vision, Raytracer Pty. Ltd, Victoria, Australia. URL: http://www. povray. org* (2004).

37  Kresse, G. & Furthmuller, J. Efficient iterative schemes for ab initio total-energy calculations using a plane-wave basis set. *Phys Rev B* **54**, 11169-11186, doi: 10.1103/PhysRevB.54.11169 (1996).

38  Perdew, J. P., Burke, K. & Ernzerhof, M. Generalized gradient approximation made simple. *Phys Rev Lett* **77**, 3865-3868, doi: 10.1103/PhysRevLett.77.3865 (1996).

39  Pack, J. D. & Monkhorst, H. J. Special Points for Brillouin-Zone Integrations - Reply. *Phys Rev B* **16**, 1748-1749, doi: 10.1103/PhysRevB.16.1748 (1977).

40  Wang, L. H. *et al.* New twinning route in face-centered cubic nanocrystalline metals. *Nat Commun* **8**, doi:10.1038/s41467-017-02393-4 (2017).

41  Asaro, R. J., Krysl, P. & Kad, B. Deformation mechanism transitions in nanoscale fcc metals. *Philosophical Magazine Letters* **83**, 733-743, doi:10.1080/09500830310001614540 (2003).

42  Chen, M. W. *et al.* Deformation twinning in nanocrystalline aluminum. *Science* **300**, 1275-

1277, doi:10.1126/science.1083727 (2003).

43   Lagerlof, K. P. D., Castaing, J., Pirouz, P. & Heuer, A. H. Nucleation and growth of deformation twins: a perspective based on the double-cross-slip mechanism of deformation twinning. *Philos Mag A* **82**, 2841-2854, doi:10.1080/01418610210157931 (2002).

44   Zhu, T. & Li, J. Ultra-strength materials. *Prog Mater Sci* **55**, 710-757, doi:10.1016/j.pmatsci.2010.04.001 (2010).

45   Bacon, D. J., Barnett, D. M. & Scattergood, R. O. Anisotropic Continuum Theory of Lattice-Defects. *Prog Mater Sci* **23**, 51-262, doi: 10.1016/0079-6425(80)90007-9 (1978).

46   Pearson, W. B. in *A Handbook of Lattice Spacings and Structures of Metals and Alloys* Vol. 4   (ed W. B. Pearson)   131-217 (Pergamon, 1958).

47   He, Y., Schwarz, R. B., Migliori, A. & Whang, S. H. Elastic constants of single crystal γ – TiAl. *Journal of Materials Research* **10**, 1187-1195, doi:10.1557/JMR.1995.1187 (2011).

**Acknowledgements**

This research was supported by the National Research Foundation of Korea (NRF) (2016R1C1B2011979, 2016M3D1A1900038, 2016R1D1A1B03932734 and 2015M1A2A2056561)

**Author contributions Statement**

B.J. and S.R. designed the research, interpret the results, and wrote the manuscript. B.J., J.K., and T.L. carried out computer simulations. S.K. discussed the results.

**Additional information**

The authors declare no competing interests.

**Figures**

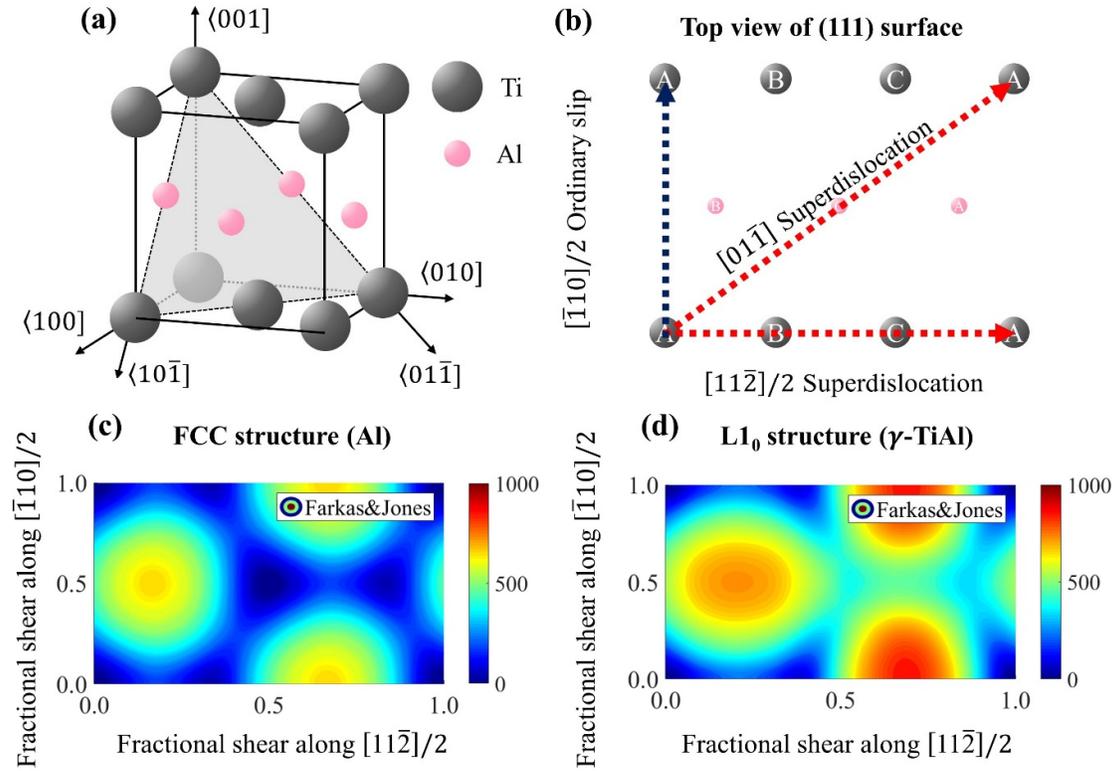

**Fig. 1.** (a) Atomic configuration of γ-TiAl having $L1_0$ crystal structure. (b) Schematic diagram of deformations, such as ordinary slip or superdislocations. (c) and (d) show the GSFE landscape for the FCC structure (Al) and the γ-TiAl single crystal, respectively.

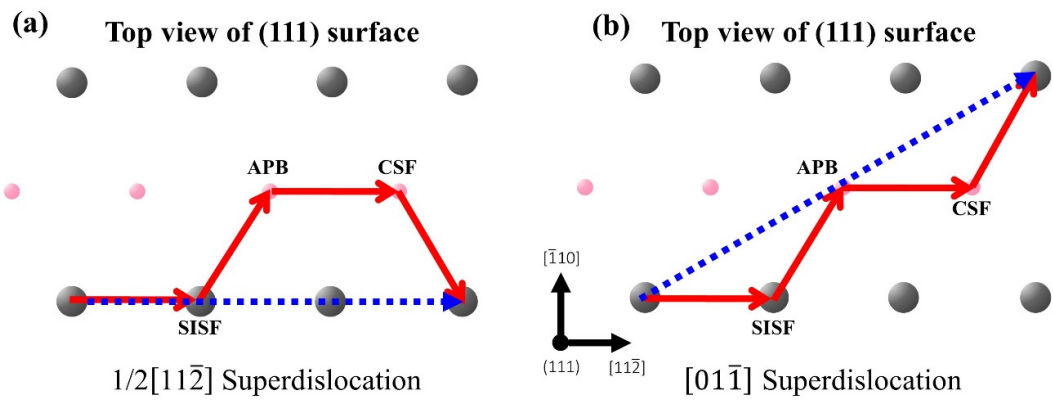

**Fig. 2.** Dissociation of (a) $1/2[11\bar{2}]$ superdislocation and (b) $[01\bar{1}]$ superdislocation into four partials.

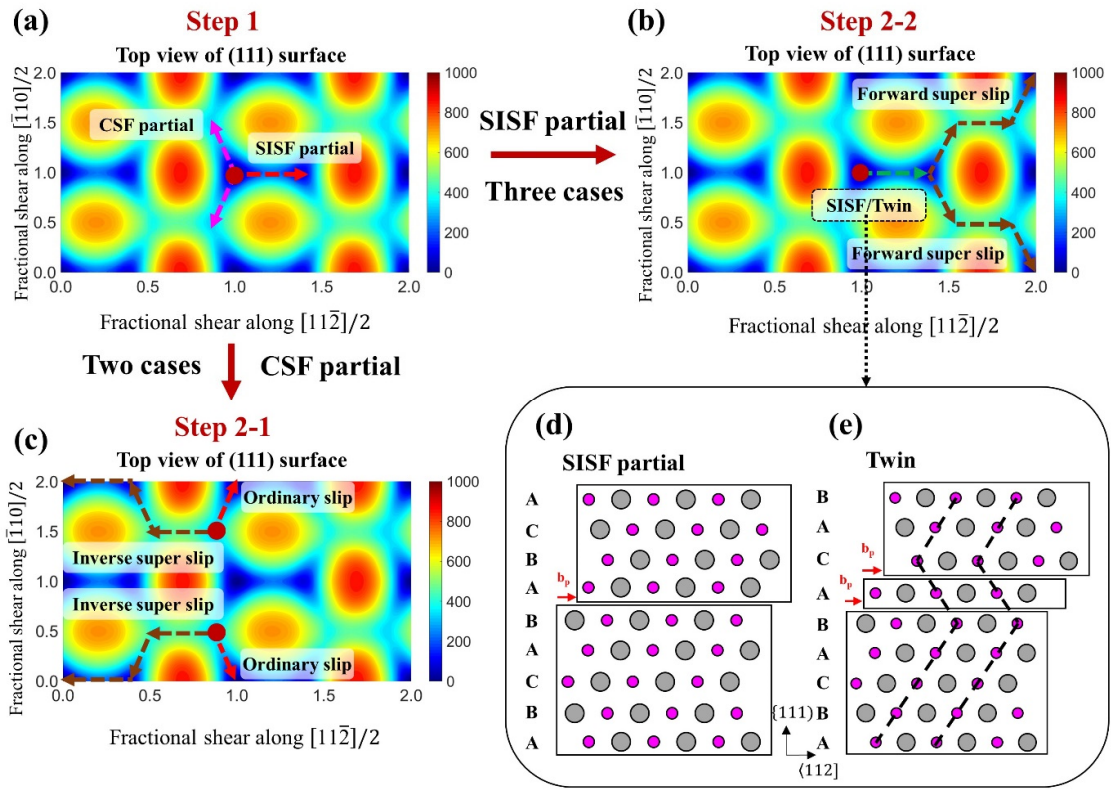

**Fig. 3.** Schematic diagram for the process of theoretical analysis of a γ-TiAl single-crystal nanowire.

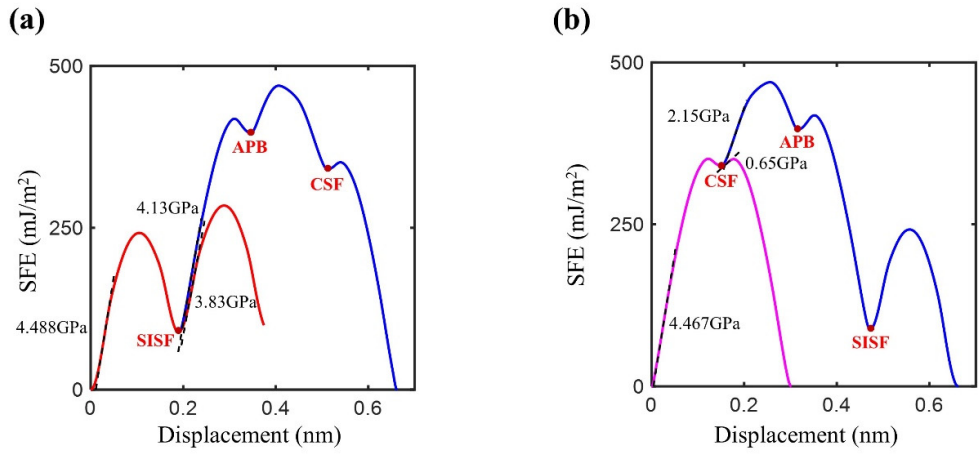

**Fig. 4.** GSFE curve when (a) SISF partial is preferred and (b) CSF partial is preferred. We compute the ideal critical resolved shear stress $\sigma^{ICRSS}$ of each partial slip from the maximum slope of the corresponding section of the GSFE curve.

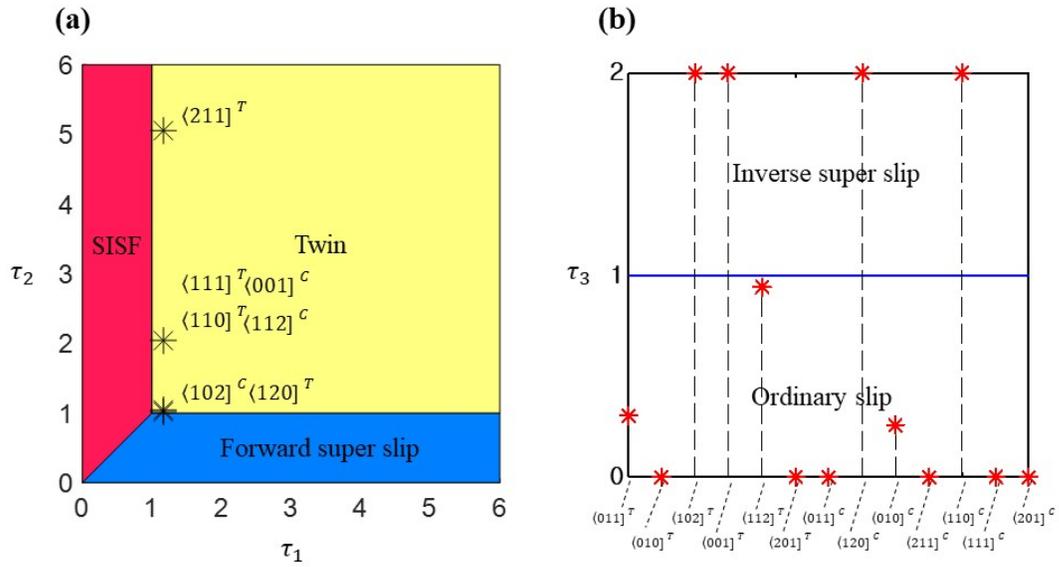

**Fig. 5.** (a) Schematic diagram showing the preferred deformation mode among multiple SISF partial, twin and forward super slip. (b) Preferred deformation mode between inverse super slip or ordinary slip when CSF partial is preferred. In the figure, $\tau_1 = \sigma_{SISF}^C/\sigma_{twin}^C$, $\tau_2 = \sigma_{Fsuper}^C/\sigma_{twin}^C$ and $\tau_3=\sigma_o^c/\sigma_{Isuper}^c$.

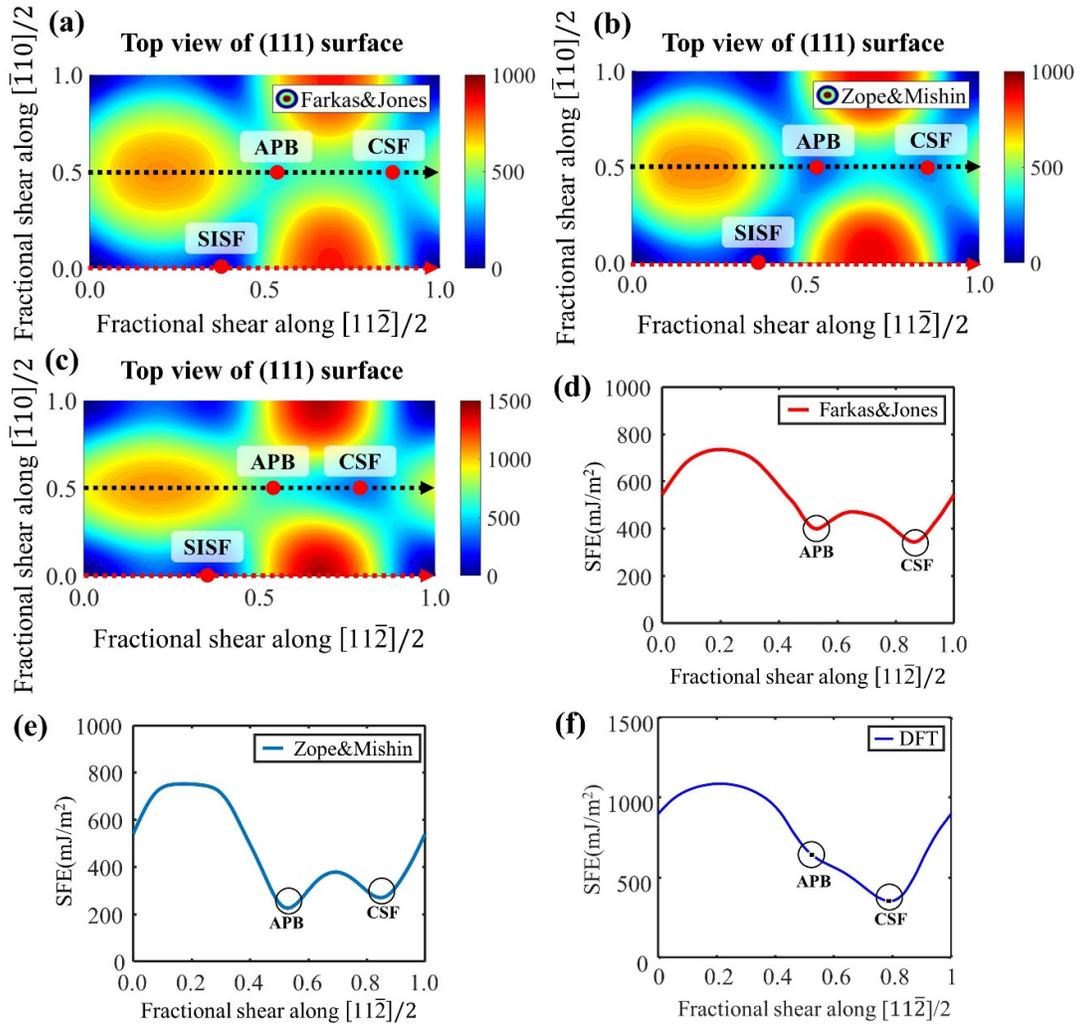

**Fig. 6.** 2-D generalized stacking fault energy profile of (a) Farkas & Jones EAM potential, (b) Zope & Mishin EAM potential and (c) DFT calculation. Generalized stacking fault energy curve showing the APB and CSF energies of (d) Farkas & Jones EAM potential and (e) Zope & Mishin EAM potential. (f) Generalized stacking fault energy calculation using density function theory (DFT) indicating that the APB energy is higher than the CSF energy.

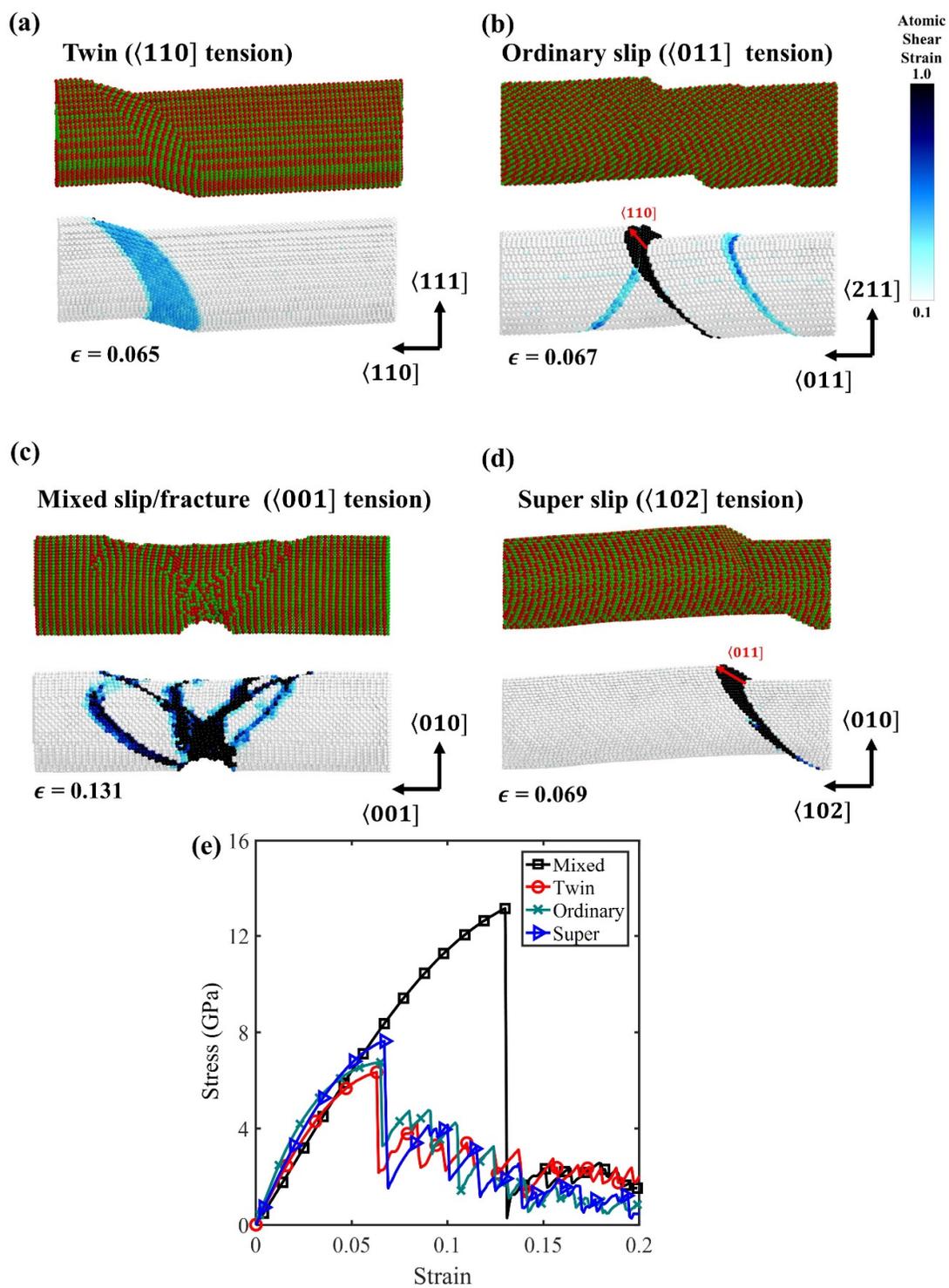

**Fig. 7.** (a) Four major deformation mechanisms performed by molecular dynamics simulations: (a) twin, (b) ordinary slip (c) mixed slip/fracture and (d) super slip. (e) Stress-strain curve for four deformation modes of (a), (b), (c) and (d).

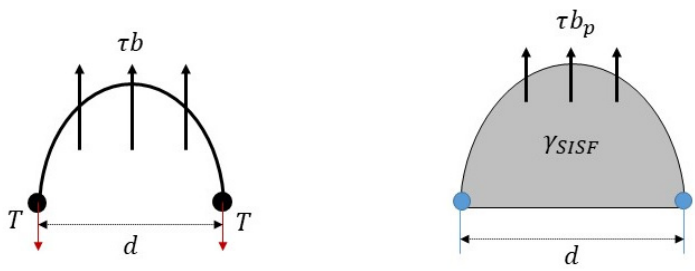

**Fig. 8.** Generation of superdislocation and twinning partial from the Frank-Read source.

**Table 1.** The lattice constant, elastic constants, stacking fault energies and effective shear modulus obtained from two EAM potentials, DFT and experiments. The effective shear modulus of the experiments is calculated based on the elastic constants determined from experiments.

|  | Farkas & Jones (Farkas and Jones, 1996) | Zope & Mishin (Zope and Mishin, 2003) | DFT | Experiment |
|---|---|---|---|---|
| a (Å) | 3.906 | 3.998 | 3.980 | 3.997[a] |
| c/a | 1.063 | 1.047 | 1.025 | 1.02 [a] |
| C11 (GPa) | 245 | 196 | 163 | 186[b] |
| C12 (GPa) | 116 | 107 | 96 | 72[b] |
| C13 (GPa) | 191 | 114 | 89 | 74[b] |
| C33 (GPa) | 352 | 213 | 156 | 176[b] |
| C44 (GPa) | 146 | 92 | 106 | 101[b] |
| C66 (GPa) | 71 | 85 | 65 | 77[b] |
| SISF (mJ/m$^2$) | 87.51 | 57.63 | 179.1 | 143[c] |
| APB (mJ/m$^2$) | 396.2 | 214.5 | 640.9 | 253[c] |
| CSF (mJ/m$^2$) | 340.7 | 260.9 | 352.1 |  |
| $\mu_{SISF}$ | 85.01 | 64.32 | 61.39 | 73.77 |
| $\mu_{APB}$ | 72.05 | 64.00 | 53.76 | 68.94 |
| $\mu_{CSF}$ | 80.92 | 60.78 | 55.55 | 70.58 |

[a]Pearson, (1958)[46]

[b]He et al., (1995)[47]

[c]Hug et al., (1988)[25]

**Table 2.** The maximum slope of SISF partial, twin and CSF partial and trailing partials of superdislocation from the Farkas & Jones EAM potential (Farkas and Jones, 1996).

| | | | |
|---|---|---|---|
| SISF partial | 4.488 GPa | Forward super slip | 4.129 GPa |
| Twin | 3.83 GPa | | |
| CSF partial | 4.467 GPa | Inverse super slip | 2.148 GPa |
| | | Ordinary slip | 0.651 GPa |

**Table 3**. Prediction based on GSFE from the Farkas & Jones EAM potential (Farkas and Jones, 1996) and the MD simulation results of 10 different orientations deformed by tension or compression. Critical stress of deformation for each loading direction are also summarized.

| Direction | Step 1 | | Step 2-1 (CSF is preferred) | | Step 2-2 (SISF is preferred) | | | $\sigma^c$ (GPa) | Prediction | Simulation |
|---|---|---|---|---|---|---|---|---|---|---|
| | $\sigma^c_{CSF}$ (GPa) | $\sigma^c_{SISF}$ (GPa) | $\sigma^c_{Isuper}$ (GPa) | $\sigma^c_o$ (GPa) | $\sigma^c_{SISF}$ (GPa) | $\sigma^c_{twin}$ (GPa) | $\sigma^c_{Fsuper}$ (GPa) | | | |
| ⟨011⟩T | 9.43 | 210.12 | 8.53 | 2.59 | | | | 9.43 | Ordinary slip | Ordinary slip |
| ⟨120⟩T | 21.85 | 10.81 | | | 10.81 | 9.23 | 9.92 | 10.81 | Twin | Twin |
| ⟨010⟩T | 16.33 | 19.46 | INF | 1.37 | | | | 16.33 | Ordinary slip | Ordinary slip |
| ⟨211⟩T | 11.18 | 11.06 | | | 11.06 | 9.45 | 62.78 | 11.06 | Twin | Twin |
| ⟨110⟩T | INF | 9.73 | | | 9.73 | 8.31 | 18.95 | 9.73 | Twin | Twin |
| ⟨102⟩T | 11.01 | INF | 4.97 | 24.35 | | | | 11.01 | Inverse super slip | Inverse super slip |
| ⟨001⟩T | 21.96 | INF | 4.66 | INF | | | | 21.96 | Inverse super slip | Mixed slip/ fracture |
| ⟨112⟩T | 12.08 | 157.59 | 6.99 | 6.49 | | | | 12.08 | Ordinary slip | Ordinary slip |
| ⟨111⟩T | 14.79 | 13.36 | | | 13.36 | 11.41 | 26.01 | 13.36 | Twin | Twin |
| ⟨201⟩T | 9.99 | 22.27 | 125.71 | 1.50 | | | | 9.99 | Ordinary slip | Ordinary slip |
| ⟨011⟩C | 17.76 | 17.81 | INF | 1.37 | | | | 17.76 | Ordinary slip | Ordinary slip |
| ⟨120⟩C | 10.97 | INF | 5.18 | 15.11 | | | | 10.97 | Inverse super slip | Inverse super slip |
| ⟨010⟩C | 9.40 | INF | 9.32 | 2.38 | | | | 9.40 | Ordinary slip | Ordinary slip |
| ⟨211⟩C | 13.68 | 53.44 | INF | 1.63 | | | | 13.68 | Ordinary slip | Ordinary slip |
| ⟨110⟩C | 21.95 | INF | 4.66 | INF | | | | 21.95 | Inverse super slip | Mixed slip/ fracture |
| ⟨102⟩C | 167.08 | 10.39 | | | 10.39 | 8.87 | 9.93 | 10.39 | Twin | Twin |
| ⟨001⟩C | INF | 9.73 | | | 9.73 | 8.31 | 18.95 | 9.73 | Twin | Twin |
| ⟨112⟩C | 44.53 | 13.36 | | | 13.36 | 11.41 | 26.01 | 13.36 | Twin | Twin |
| ⟨111⟩C | 26.66 | 29.20 | INF | 2.15 | | | | 26.66 | Ordinary slip | Ordinary slip |

| 〈201]C | 10.32 | 262.65 | INF | 1.46 | | 10.32 | Ordinary slip | Ordinary slip |

* Infinite critical stress indicates that the Schmid factor for the deformation mode is negative or 0.

**Table 4.** Ideal critical resolved shear stress of each deformation mode from DFT.

| | | | |
|---|---|---|---|
| SISF partial | 4.974 GPa | Forward super slip | 7.859 GPa |
| Twin | 4.586 GPa | | |
| CSF partial | 7.469 GPa | Inverse super slip | 4.856 GPa |
| | | Ordinary slip | 4.228 GPa |

**Table 5.** Comparison of experiments (Feng and Whang, 1999; Inui et al., 1997; Stucke et al., 1995) and the theoretical prediction based on GSFE from DFT.

| Direction | Step 1 | | Step 2-1 (CSF is preferred) | | Step 2-2 (SISF is preferred) | | | $\sigma^c$ (GPa) | Experiment | Prediction (DFT) |
|---|---|---|---|---|---|---|---|---|---|---|
| | $\sigma^c_{CSF}$ | $\sigma^c_{SISF}$ | $\sigma^c_{Isuper}$ | $\sigma^c_o$ | $\sigma^c_{SISF}$ | $\sigma^c_{twin}$ | $\sigma^c_{Fsuper}$ | | | |
| $[001]^C$ (RT) | INF | 10.64 | | | 10.64 | 9.81 | 39.89 | 10.64 | Super slip | Twin |
| $[\bar{1}10]^C$ (RT) | 26.11 | INF | 10.39 | INF | | | | 26.11 | Super slip | Inverse super slip |
| $[\bar{1}52]^C$ (RT) | 17.91 | 1073 | 41.3 | 15.48 | | | | 17.91 | Super slip | Ordinary slip |
| $[021]^C$ (RT) | 19.49 | 715.2 | INF | 10.79 | | | | 19.49 | Ordinary slip | Ordinary slip |
| $[\bar{2}51]^C$ (RT) | 16.38 | INF | 11.43 | INF | | | | 16.38 | Super slip | Inverse super slip |
| $[\bar{2}33]^C$ (RT) | 41.28 | 23.15 | | | 23.15 | 21.35 | INF | 23.15 | Super slip | Twin |
| $[\bar{1}91]^C$ (RT) | 15.87 | INF | 15.86 | 44.15 | | | | 15.87 | Super slip | Inverse super slip |
| $[\bar{1}63]^C$ (873K) | 19.79 | 107.4 | 113.28 | 13.01 | | | | 19.79 | Ordinary slip | Ordinary slip |
| $[\bar{1}\,12\,5]^C$ (873K) | 17.87 | INF | 71.85 | 12.61 | | | | 17.87 | Ordinary slip | Ordinary slip |

\* Infinite critical stress indicates that the Schmid factor for the deformation mode is negative or 0.

# Supplementary Information for

# Systematic investigation of the deformation mechanisms of a γ-TiAl single crystal

Byungkwan Jeong[a], Jaemin Kim[a], Taegu Lee[a], Seong-Woong Kim[b,**], Seunghwa Ryu[a,*]

[a] Department of Mechanical Engineering & KI for the Nano Century, Korea Advanced Institute of Science and Technology, Daejeon 34141, Republic of Korea

[b] Titanium Department, Korea Institute of Materials Science, Changwon 51508, Republic of Korea

* Corresponding author. Tel.: +82-42-350-3019; fax: +82-42-350-3059

Email address: ryush@kaist.ac.kr (Seunghwa Ryu)

** Co-corresponding author. Tel.: +82-55-280-3000; fax: +82-55-280-3333

Email address: mrbass@kims.re.kr (Seong-Woong Kim)

**Supplementary Note 1.** The ideal critical resolved shear stress to emit a twinning partial.

Twinning partial involves stacking fault energy unlike full dislocation. Therefore, in order to analyze the twinning partial emission, we must consider the effect of the stacking fault energy. In order to calculate the ICRSS of a twinning partial, we need to consider Peierls stress, the interaction force between leading and trailing partial and stacking fault energy. Therefore, the critical resolved shear stress to emit a twinning partial is given as [1],

$$\tau_{tp} = \frac{\mu_{tp} b_{tp}}{d} + \frac{\gamma_{SISF} - F_{12}}{b_{tp}} + \tau_p \quad (1)$$

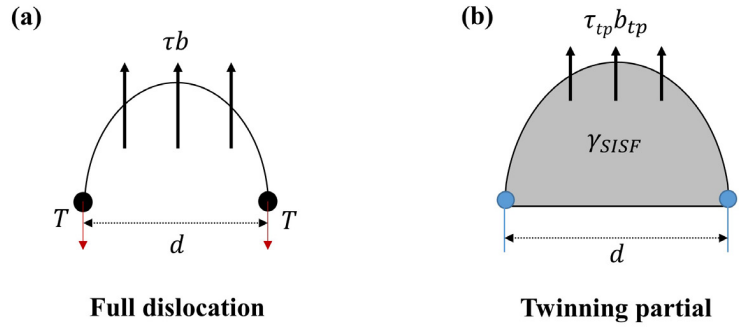

**Figure S1.** Comparison of the ideal critical resolved shear stress of full dislocation and twinning partial.

where $\tau_{tp}$ is the ideal critical resolved shear stress to emit a twinning partial, $\gamma_{SISF}$ is the stacking fault energy, $d$ is the sample size, $b_{tp}$ is the magnitude of the Burgers vector of the twinning partial, $\mu_{tp}$ is the effective shear modulus for the twinning partial, $F_{12}$ is the interaction force between the leading and trailing partial, and $\tau_p$ is the Peierls stress. The lattice friction stress (the Peierls stress) at 0K is estimated by a DFT method to be small value of around 0.01C44 (C44=68 GPa) [2]. The lattice friction stress at 300K is significantly smaller due to thermal fluctuation as shown earlier studies (Kang et al., 2014). Furthermore, there is no interaction between leading and trailing partial when twinning is observed, and we can approximately simplify (1) as follows.

$$\tau_{tp} \approx \frac{\mu_{tp} b_{tp}}{d} + \frac{\gamma_{SISF}}{b_{tp}} \quad (2)$$

**Supplementary Note 2.** Calculation of the effective shear modulus

To calculate the effective shear modulus, we first calculated elastic constants for γ-TiAl single crystal using Farkas & Jones EAM potential and using first principle calculation. Following the method of Scattergood and Bacon [3], the effective shear modulus can be calculated as follows:

$$\mu = \frac{4\pi}{b^2} E_s \qquad (6)$$

where $E_s$ is the pre-logarithmic energy factor of screw dislocations. All the results of the elastic constants and the effective shear modulus can be found in Table 1.

**Supplementary Figure**

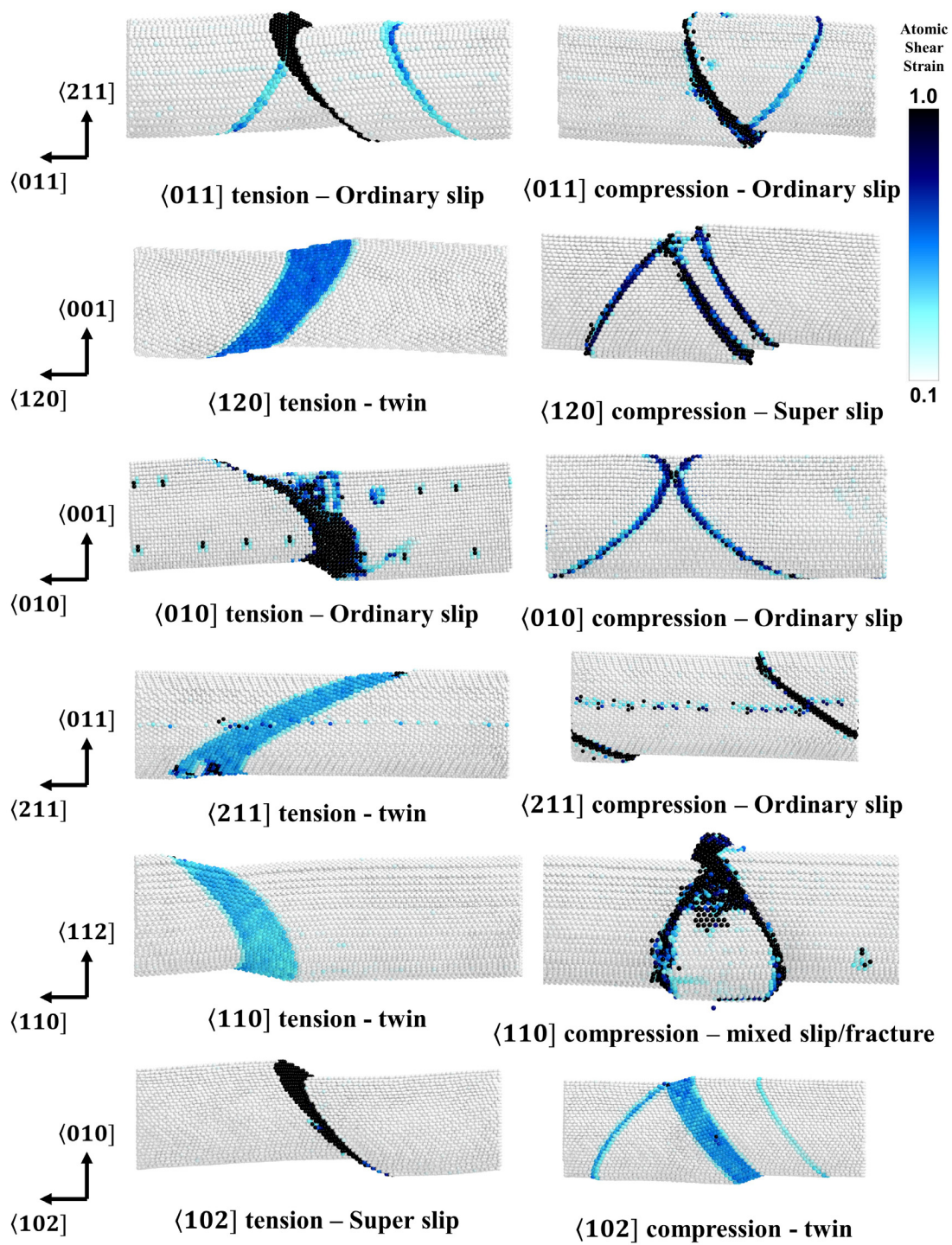

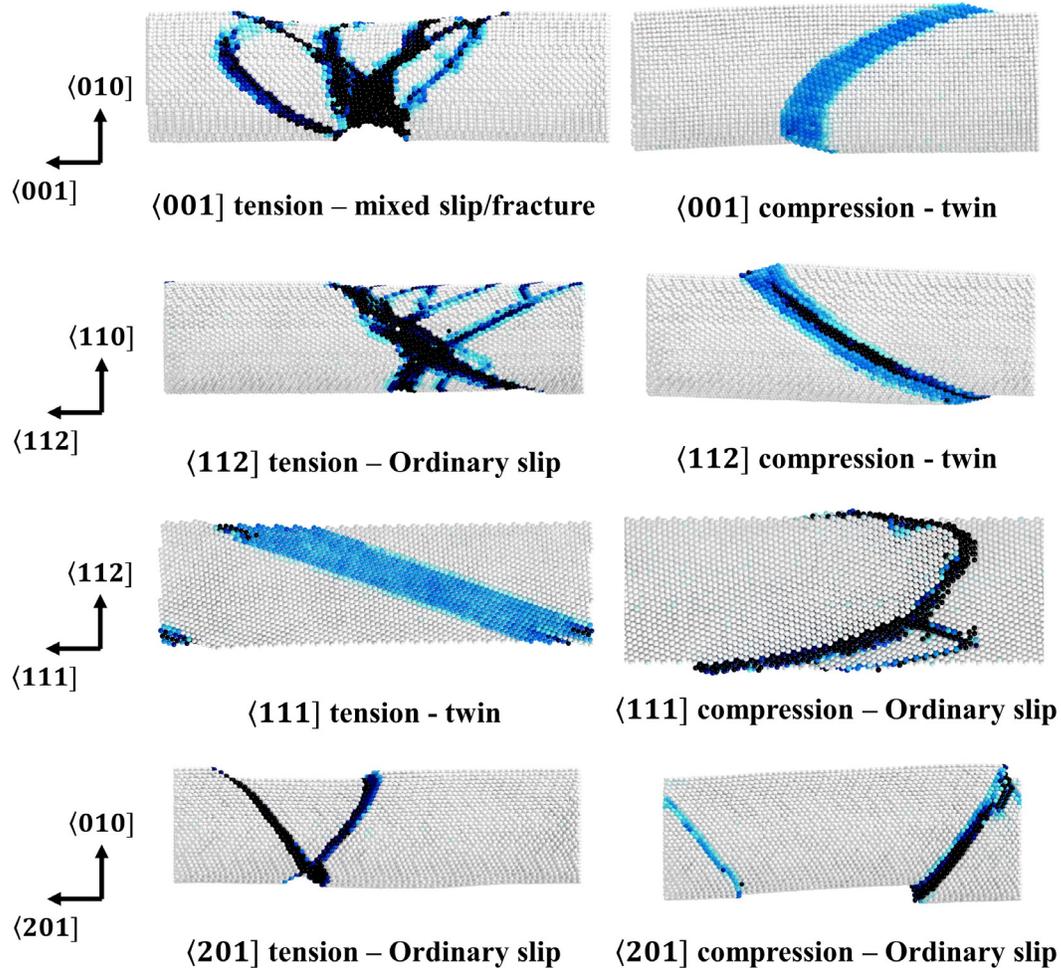

**Fig. S2.** Simulation results for 10 orientations under uniaxial tension/compression.

Fig. S2 shows deformation results for 10 different orientations under uniaxial tension or compression. For each orientation, the observed deformation modes are also indicated.

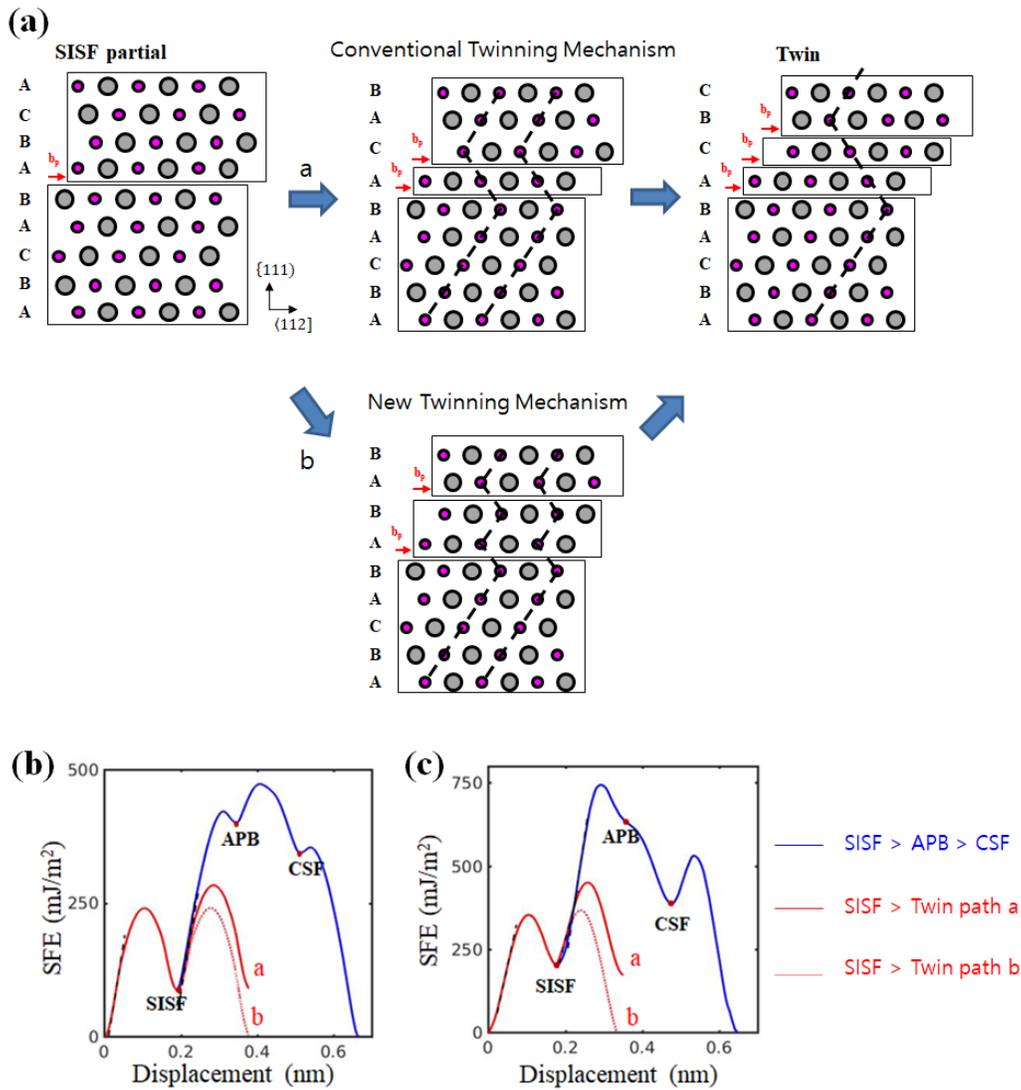

**Fig. S3.** (a) Schematic of conventional and new twinning path reported by Wang et al. [5] (b) Stacking fault energy curve obtained from Farkas EAM potential (c) Stacking fault energy curve obtained from DFT calculation

Fig. S3 shows the schematic of conventional and new twinning path reported by Wang et al.[5], as well as the generalized stacking fault energy curves obtained by Farkas EAM potential and DFT calculations. For both calculations (Fig S3(b)-(c)), it is evident that the new twinning path has lower ideal critical resolved shear stress (ICRSS) than the conventional path does. Still, whichever twining path is considered, twinning is predicted to be the preferred deformation mechanism along [001] and [$\bar{2}$33] compressions, while experiments show that samples under these two loading conditions are deformed by superdislocations.

**Supplementary Table S1.** Critical stress of partial slips for the [$\bar{1}52$] orientation in compression loading on {111} slip planes.

| {111} slip planes | Step 1 | | Step 2 | | Prediction |
|---|---|---|---|---|---|
| | $\sigma^c_{CSF}$ (GPa) | $\sigma^c_{SISF}$ (GPa) | $\sigma^c_o$ (GPa) | $\sigma^c_{Isuper}$ (GPa) | |
| (111) | 19.42 | 1073 | 10.83 | INF | Ordinary slip |
| (11$\bar{1}$) | 35.15 | INF | 77.91 | 37.6 | Inverse super slip |
| (1$\bar{1}$1) | 20.15 | INF | INF | 15.3 | Inverse super slip |
| ($\bar{1}$11) | 17.91 | INF | 15.48 | 41.3 | Ordinary slip |

For [$\bar{1}52$] orientation, which is deformed by super slip on the (1$\bar{1}$1) slip plane in the experiment [4], we predicted that the preferred deformation mode is ordinary slip on the ($\bar{1}$11) slip plane because it has the lowest $\sigma^c_{CSF}$. However, $\sigma^c_{CSF}$ on ($\bar{1}$11) and (1$\bar{1}$1) slip planes differ by approximately 11%, which can be caused by different Al composition or the limitation of the accuracy of the DFT calculation.

# Reference

[1] K.P.D. Lagerlof, J. Castaing, P. Pirouz, A.H. Heuer, Nucleation and growth of deformation twins: a perspective based on the double-cross-slip mechanism of deformation twinning, Philos. Mag. A. 82 (2002) 2841–2854. doi:10.1080/01418610210157931.

[2] C. Woodward, S.I. Rao, Ab-initio simulation of (a/2)<110] screw dislocations in γ-TiAl, Philos. Mag. 84 (2004) 401–413. doi:10.1080/14786430310001611626.

[3] D.J. Bacon, D.M. Barnett, R.O. Scattergood, Anisotropic continuum theory of lattice defects, 23 (1979).

[4] H. Inui, M. Matsumuro, D.-H. Wu, M. Yamaguchi, Temperature dependence of yield stress, deformation mode and deformation structure in single crystals of TiAl (Ti−56 at.% Al), Philos. Mag. A. 75 (1997) 395–423. doi:10.1080/01418619708205149.

[5] L. Wang, P. Guan, J. Teng, P. Liu, D. Chen, W. Xie, D. Kong, S. Zhang, T. Zhu, Z. Zhang, E. Ma, M. Chen, X. Han, New twinning route in face-centered cubic nanocrystalline metals, Nat. Commun. 8 (2017) 2142. doi: 10.1038/s41467-017-02393-4.